\documentclass[11pt,A4]{article}
\usepackage{graphics}
\begin{document}

\begin{center}
{\large \bf INVESTIGATION OF DILUTE MAGNETIC SYSTEMS WITH SPIN-1 ISING MODEL IN THE FRAME OF GENERALIZED STATISTICAL MECHANICS}\\

\vspace{0.7cm}
{\large Metin Karabekirogullari$^{*1,2}$, Fevzi B\"{u}y\"{u}kk\i l\i \c{c}$^{**1}$,  Dogan Demirhan}$^{***1}$\\
$^{1}$Ege University, Faculty of Science, Department of
Physics, Izmir-TURKEY.\\
$^{2}$Pamukkale University,Faculty of Arts and
Sciences,Department of Physics, Denizli-TURKEY.\\

\end{center}


\begin{abstract}
In this study the magnetization phenomenon has been investigated
as a behavior of interacting elementary moments ensemble, with the
help of Ising model [1] in the frame of non-extensive statistical
mechanics. To investigate the physical systems with three states
and two order parameters, the spin-1 single lattice Ising model or
three states systems are used. In the manner of this model
thermodynamical properties of a great deal of physical phenomena
such as ferromagnetism in bilateral alloys, liquid mixtures,
liquid-crystal mixtures, freezing, magnetic orderliness, phase
transformations, semi-stable and unstable states, ordered and
disordered transitions [2,3,4,5].

\vspace{0.7 cm} {\small Keywords: Non-extensive statistical
mechanics}


\vspace{.7cm} {\small $^*$ Corresponding
Authors:e-mail:metink@sci.ege.edu.tr, Phone:+90 232-3881892
(ext.2363)
Fax:+90 232-3881892\\
\\
} \vspace{.7cm} {\small $^{**}$ e-mail:fevzi@sci.ege.edu.tr,
Phone:+90 232-3881892 (ext.2846)
Fax:+90 232-3881892\\
\\
} \vspace{.7cm} {\small $^{***}$ e-mail:dogan@sci.ege.edu.tr,
Phone:+90 232-3881892 (ext.2381)
Fax:+90 232-3881892\\
\\
}
\end{abstract}

\section{Introduction}

In this study, taking spin-1 Ising systems as the model dilute
magnetic systems have been investigated in the  frame of
non-extensive statistics [6,7,8,9]. The general Hamiltonian of
next nearest pair interaction spin-1 Ising systems is in the form

\begin{eqnarray}
   \emph{H}=-J\sum_{<ij>}S_{i}S_{j}-K\sum_{<ij>}S_{i}^{2}S_{j}^{2}- \nonumber \\
   D\sum_{j=1}^{N}S_{j}^{2}-H\sum_{j=1}^{N}-L\sum_{<ij>}S_{i}Q_{j}- \nonumber \\
   M\sum_{<ij>}S_{i}S^{j}(S_{i}+S_{j}) \label{Eq:1}
\end{eqnarray}
where J, K, D, H, L, M and N are bilinear exchange interaction
constant, bi-quadratic exchange interaction constant, crystal
field interaction constant, magnetic  field due to S',
dipole-quadropole interaction constant, magnetic perturbation of
third degree and number of lattice points respectively. It the
systems stay in semi-stable state or phase their properties change
considerably. For examples, some alloys or metals lead to
amorphons (glassy metal) structure when they are rapidly cooled
down. This structure namely the phase is a semi-stable state. In
this manner, some of the characteristics of the alloys or metals
such as magnetic properties, resistance against corrosion,
exhausting, wearing and hardness improve considerably. This
Hamiltonian includes all possible terms when $s_{i}^{\beta}=s_{i}$
it does not include higher order powers of spin.

\section{Dilute Systems}\label{Intro}
Dilute magnetic system is a system which is obtained by inclusion
of non-magnetic atoms to magnetic atoms. Exhibition of magnetic
property of such systems is possible when the concentration of the
magnetic atoms reaches to a certain value. Dilute magnetism is
given by the expression $A_{p}B_{1-p}C$ where A shows magnetic, B
non-magnetic and C purely magnetic states. The Hamiltonian of such
a system is defined in the form

\begin{equation}\label{Eq:2}
    \emph{H}=-\sum_{<ij>}S_{i}S_{j}Jn_{i}n_{j}
\end{equation}

which is obtained by taking M=K=L=0 in Eq.(1). In this equation
$S_{i}$ and $s_{j}$ are the spin vectors of i th and j th atoms,
$n_{i}$ and $n_{j}$ are spin disorder variables taking the values
0 and 1 whose average value gives the magnetic concentration c. J
is exchange energy.

\begin{table}[h]
\caption{ Spin Shaping Probabilities.}
\label{tab:1}       
\begin{tabular}{llllll}
\hline\noalign{\smallskip}
Shaping of Pairs & Probability & Energy & $\mu_i$ & Spin & Probability   \\
\noalign{\smallskip}\hline\noalign{\smallskip}
1-1  & $y_{11}$ & $-(\frac{J}{2})+\epsilon_{11}$ & 1 & 1 & $x_1$ \\
1-2  & $y_{12}$ & $\epsilon_{12}$ &1 & 2 & $x_2$ \\
1-3  & $y_{13}$ & $+(\frac{J}{2})+\epsilon_{13}$ & 1 & 3 & $x_3$ \\
2-1  & $y_{21}$ & $\epsilon_{21}$ & 1 &   &   \\
2-2  & $y_{22}$ & $\epsilon_{22}$ & 1 &   &   \\
2-3  & $y_{23}$ & $\epsilon_{23}$ & 1 &   &   \\
3-1  & $y_{31}$ & $+(\frac{J}{2})+\epsilon_{31}$ & 1 &   &   \\
3-2  & $y_{32}$ & $\epsilon_{32}$ & 1 &   &   \\
3-3  & $y_{33}$ & $-(\frac{J}{2})+\epsilon_{33}$ & 1 &   &   \\
\noalign{\smallskip}\hline \\
\end{tabular}
\vspace*{1.0cm}  
\end{table}

In table 1 where the spin shaping probabilities are presented,
$\nu_{i}$ are different shping numbers having same probability.
$\epsilon_{11}$, $\epsilon_{12}$, $\epsilon_{13}$,
$\epsilon_{21}$, $\epsilon_{22}$, $\epsilon_{23}$,
$\epsilon_{31}$, $\epsilon_{32}$ and $\epsilon_{33}$ are atomic
interactions. If magnetic atoms are grouped as A and B denotes the
atoms which are not magnetic then
$\epsilon_{11}=\epsilon_{13}=\epsilon_{33}=\epsilon_{AA},
\epsilon_{12}=\epsilon_{32}=
\epsilon_{21}=\epsilon_{23}=\epsilon_{AB} and
\epsilon_{22}=\epsilon_{BB}$. The spin directions of the magnetic
atoms are 1 and 3 and the directions of the atoms which are not
magnetic is 2. The probabilities of the atoms having spins in the
upward directions is $x_{1}$, ,n downward is $x_{3}$ and the atoms
that are not magnetic is $x_{2}$. Since the probabilities having
symmetry are equal as a result $y_{12}=y_{21}, y_{23}=y_{32}$ and
$y_{13}=y_{31}$. The shaping of pairs at the lattice point takes
place in nine different form and these shaping have been given in
Table 1.

\section{Free Energy of the System}

The energy per atom of such a system is

\begin{eqnarray}
\frac{E}{N}=J \gamma (y_{13}-\frac{y_{11}+y_{33}}{2})+
\frac{\gamma}{2} [\epsilon_{AA} (y_{11}+2 y_{13}+y_{33})+2
\epsilon_{AB}+ \nonumber \\
(y_{12}+y_{23})+\epsilon_{BB} y_{22}]+ \nonumber \\
(y_{11}+y_{12}-y_{33}-y_{32}) \beta H \label{Eq:3}
\end{eqnarray}

where $\gamma$ is the number of nearest lattice points, N is the
number of lattice points of the system, H is the external magnetic
field and $\beta=\frac{1}{k_{B}T}$. Using the weight factor
$x_{i}$ as the internal variable which has been developed for
three state systems and $y_{ij}$ as double variable, the
statistical weight of the spin-1 Ising system could be written as;

\begin{equation}\label{Eq:4}
    [W]^{\frac{1}{N}}=\frac{[\prod_{i=1}^{3}(x_{i}L)!]^{\gamma-1}}{L!^{\frac{\gamma}{2}-1}[\prod_{i,j=1}^{3}(y_{ij}L)!]^{\frac{\gamma}{2}}}
\end{equation}

where N is the number of lattice points in the system and 1 is the
number of systems in the ensemble. On the other hand the
definition of entropy is

\begin{equation}\label{Eq:5}
    S=\frac{k_B}{lnW}.
\end{equation}

When the entropy is calculated for a single system (L=1)

\begin{equation}\label{Eq:6}
\frac{S}{N}=k_B[(\gamma-1) \sum_{i=1}^3 x_i
lnx_i-(\frac{\gamma}{2})\sum_{i,j=1}^3 y_{ij} lny_{ij}]
\end{equation}

is found. Thus free energy becomes:

\begin{eqnarray}
\nonumber
 F=E-TS .
\end{eqnarray}

Using the expressions in Table 1, free energy takes the form

\begin{eqnarray}
F=J \gamma (y_{13}-\frac{y_{11}+y_{33}}{2})+ \nonumber \\
\frac{\gamma}{2} [\epsilon_{AA} (y_{11}+2 y_{13}+y_{33})+2
\epsilon_{AB} (y_{12}+y_{23})+\epsilon_{BB}
y_{22}]- \nonumber \\
(y_{11}+y_{12}-y_{33}-y_{32}) \beta H-\frac{1}{\beta}[(\gamma-1)
\sum_{i=1}^3 x_i lnx_i-(\frac{\gamma}{2})\sum_{i,j=1}^3 y_{ij}
lny_{ij}] \label{Eq:7}
\end{eqnarray}

\section{Obtaining the Equilibrium State of the Dilute System}
The equilibrium state of a dilute system could be obtained by the
calculus of variations. For this purpose three independent
variables $\eta$, $\xi_{1}$ and $\xi_{2}$ are defined:

\begin{equation}\label{Eq:8}
    \eta=\frac{y_{13}+y_{31}}{2}=y_{13}
\end{equation}

\begin{equation}\label{Eq:9}
    2\xi_{1}=x_{1}-x_{2}
\end{equation}

\begin{equation}\label{Eq:10}
    2\xi_{2}=y_{12}-y_{32}
\end{equation}

where $\eta$ indicates the energy variation, $\xi_{1}$ represents
magnetization and $\xi_{2}$ shows the abundance of the upward
spins with respect to the downward spins of A atoms which are
neighbors of B atoms.

\section{Equilibrium State of the Dilute System in the Frame of Non-extensive Statistical Mechanics}
Until 1998 all of the physical quantities of statistical systems
were obtained by Boltzmann-Gibbs statistics. According to
Boltzmann-Gibbs statistics, macro quantities such as free energy,
entropy and internal energy of a statistical system were accepted
as extensive quantities. In 1998 a generalization in the
Thermodynamical meaning has been carried aut to understand the
structure or to solve a great number of unfamiliar systems and the
generalization was inspired from the probability definition of the
multifractal geometry. Magnetization is a process with a long
range and memory. Such systems are non-extensive. Thus
magnetization will be investigated in the frame of non-extensive
statistical mechanics. This generalization is the parametrization
of all of the statistical quantities by a parameter q. In the
limit $q\rightarrow1$ the statistics under investigation reduces
to Boltzmann-Gibbs statistics. For the values of q different than
1, macro quantities such as internal energy, free energy and
entropy in Boltzmann-Gibbs statistics are not extensive quantities
in other words they are non-extensive. For generalization,
mathematical expressions obtained from non-extensive statistical
mechanics are introduced;

\begin{eqnarray}
\ln_{q}x=\frac{x^{1-q}-1}{1-q} and \nonumber \\
\exp_{q}^{x}=\{[1+(1-q)x]^{\frac{1}{1-q}} if 1+(1-q)x\geq0 \nonumber \\
    0        else . \label{Eq:11.a}
\end{eqnarray}

On the other hand, when Eq.(11a) is written down for any function
$f(x)$, its derivative becomes

\begin{equation}\label{Eq:11.b}
[ln_q f(x)=\frac{f(x)^{1-q}-1}{1-q}]^l=\frac{(1-q) f(x)^-q
f^l(x)}{1-q}=\frac{f^l(x)}{f(x)^q}
\end{equation}

In this system, the equilibrium state is determined using the
variation of the free energy with respect to parameters $\eta$,
$\xi_{1}$ and $\xi_{2}$ given above. The variation of free energy
leads to nonlinear equations:

\begin{equation}\label{Eq:12}
    y_{11} y_{33}= y_{13}^2 e_q^{\frac{2 J}{k T}}
\end{equation}

\begin{equation}\label{Eq:13}
(\frac{y_{11}}{y_{33}})^{\frac{\gamma}{2}}=(\frac{x_1}{x_3})^{\gamma-1}
e_q^{\frac{2 \beta H}{k_B T}} e_q^{(\gamma-1)
(x_3^{1-q}-x_1^{1-q})}
\end{equation}

\begin{equation}\label{Eq:14}
\frac{y_{11}}{y_{33}}= (\frac{y_{12}}{y_{32}})^2
e_q^{2(y_{12}^{1-q} - y_{32}^{1-q})+ (y_{33}^{1-q} -
y_{11}^{1-q})}
\end{equation}

Starting from Table 1 and taking $x_3^{1-q} \cong x_1^{1-q}$ in
Eq.(13) and $y_{12}^{1-q} \cong y_{32}^{1-q}$ and $y_{33}^{1-q}
\cong y_{11}^{1-q}$ in Eq.(14) these equations are

brought to simpler form:

\begin{equation}\label{Eq:15}
    y_{11} y_{33}=y_{13}^{2} e_{q}^{\frac{2J}{k_{B}T}}
\end{equation}

\begin{equation}\label{Eq:16}
(\frac{y_{11}}{y_{33}})^{\frac{\gamma}{2}}=(\frac{x_1}{x_3})^{\gamma-1}
e_q^{\frac{2 \beta H}{k_B T}}
\end{equation}

\begin{equation}\label{Eq:17}
    \frac{y_{11}}{y_{33}}=[\frac{y_{12}}{y_{32}}]^{2} .
\end{equation}

It is obvious that the solutions of these equations could be
obtained depending on the entropy parameter q. In order to proceed
to physically measurable results, let us relate the probabilities
with the relevant concentrations. Existence probability of the
magnetic atoms in the lattice point is $n_{A}$ and the probability
for the atoms that are not magnetic is $n_{B}$. Therefore

\begin{equation}\label{Eq:18}
    n_{A}=x_{1}+x_{3}
    or
    n_{A}=y_{11}+y_{12}+y_{13}+y_{31}+y_{32}+y_{33}
\end{equation}

\begin{equation}\label{Eq:19}
    n_{B}=x_{2} or n_{B}=y_{21}+y_{22}+y_{32}
\end{equation}

\begin{equation}\label{Eq:20}
    n_{AA}=y_{11}+y_{13}+y_{31}+y_{33}
\end{equation}

\begin{equation}\label{Eq:21}
    n_{AB}=y_{12}+y_{13}+y_{32}
\end{equation}

\begin{equation}\label{Eq:22}
    n_{BA}=y_{21}+y_{23}
\end{equation}

\begin{equation}\label{Eq:23}
    n_{BB}=y_{22} .
\end{equation}

Writing $c=\eta_{A}$ in other wards by taking the concentration of
the magnetic atoms above equations could be expressed in terms of
c:

\begin{equation}\label{Eq:24}
    n_{AA}=c^{2}
\end{equation}

\begin{equation}\label{Eq:25}
    n_{AB}=c(1-c)
\end{equation}

\begin{equation}\label{Eq:26}
    n_{BA}=c(1-c)
\end{equation}

\begin{equation}\label{Eq:27}
    n_{BB}=(1-c)^{2} .
\end{equation}

\section{Calculation of the Orientation Probabilities}
In order to solve the equations given above a transformation given
by

\begin{equation}\label{Eq:28}
\frac{x_1}{x_3}=e_q^{2 \gamma t}
\end{equation}

is used in Eqs.(15),(16) and (17) which leads to the solutions of
these nonlinear equations:

\begin{equation}\label{Eq:29}
    \xi_{1}=\frac{n_{A}}{2}\tanh_{q}(\gamma-1)t .
\end{equation}

The third parameter

\begin{equation}\label{Eq:30}
    \xi_{2}=\frac{n_{B}}{2}\tanh_{q}(\gamma-1)t
\end{equation}

is obtained. Using these results the probabilities $x_{i}$ and
$y_{ij}$ are found:

\begin{equation}\label{Eq:31}
    y_{11}=\frac{n_{AA}\exp_{q}^{2(\gamma-1)t}}{2[\cosh_{q}2(\gamma-1)t+e_{q}^{\frac{-J}{k_{B}T}}]}
\end{equation}

\begin{equation}\label{Eq:32}
    y_{12}=y_{21}=\frac{n_{AB}}{2}(\tanh_{q}(\gamma-1)t+1)
\end{equation}

\begin{equation}\label{Eq:33}
    y_{13}=y_{31}=\frac{n_{AA}}{2[\exp_{q}^{\frac{J}{k_{B}T}}\cosh_{q}2(\gamma-1)t+1]}
\end{equation}

\begin{equation}\label{Eq:34}
    y_{22}=n_{BB}
\end{equation}

\begin{equation}\label{Eq:35}
y_{23}=y_{32}=\frac{n_{AB}}{2}(1-\tanh_{q}(\gamma-1)t)
\end{equation}

\begin{equation}\label{Eq:36}
    y_{33}=\frac{n_{AA}}{2\exp_{q}^{\frac{J}{k_{B}T}}[\cosh_{q}2(\gamma-1})t+\exp_{q}^{\frac{-J}{k_{B}T}]}
    .
\end{equation}

On the other hand the internal variables $x_{i}$ are determined
as,

\begin{equation}\label{Eq:37}
x_1=\frac{n_A (1+\tanh_q \gamma t)}{2}
\end{equation}

\begin{equation}\label{Eq:38}
    x_{2}=n_{A}
\end{equation}

\begin{equation}\label{Eq:39}
x_3=\frac{n_A (1-\tanh_q \gamma t)}{2}
\end{equation}

\section{Determination of the Physical Quantities}
When the relation between the parameter t and temperature is
determined by using the obtained probabilities, the nonlinear
equations and expression related to the concentration one gets:

\begin{eqnarray}
e_q^{\frac{J}{kT}}=\frac{n_A \tanh_q \gamma t}{\cosh_q 2
(\gamma-1) t [n_{AA} \tanh_q 2 (\gamma-1)
t +n_{AB} \tanh_q (\gamma-1) t-n_A \tanh_q \gamma t]}- \nonumber \\
\frac{n_{AB}\tanh_q (\gamma-1) t}{\cosh_q 2 (\gamma-1) t [n_{AA}
\tanh_q 2 (\gamma-1) t +n_{AB} \tanh_q (\gamma-1) t-n_A \tanh_q
\gamma t]} \label{Eq:40}
\end{eqnarray}

According to Eq.(29) the magnetization is

\begin{equation}\label{Eq:41}
\tau_q=2\xi_1 or \tau_q=n_A\tanh(q \gamma t)
\end{equation}

Variation of the magnetization with respect to temperature at a
certain concentration value (c=1) for different q values, in other
words the plot of the function $\tau_{q}=f(T)$ is given in
Fig.(1). In the figure solid line refers to $q\cong1$, closer
spaced dashed line shows the plot for q=0.8 and wider spaced
dashed line for $q=0.5$ .

\begin{figure}[t]
\resizebox{0.45\textwidth}{!}{%
  \includegraphics{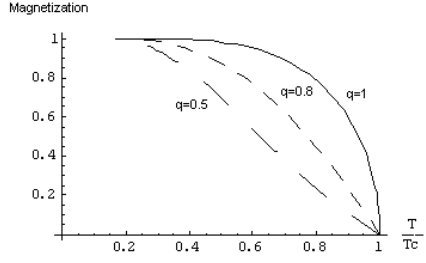}
}
\caption{Variation of magnetization with respect to temperature
for simple cubic structure ($\gamma=6$.) at a certain
concentration (c=1) and for different g values (Eq(29)).
$\tau=f(T$) plot.}
\label{fig:10}       
\end{figure}

The susceptibility is expressed as

\begin{equation}\label{Eq:42}
\frac{1}{\chi_q}=\frac{kT[1+e_q^{\frac{J}{kT}}\cosh_q 2 (\gamma-1)
t] (n_A-\gamma n_{AA}-n_{AA})+2 n_{AA}(\gamma-1)}{\beta^2
[1+e_q^{\frac{J}{kT}}\cosh_q 2 (\gamma-1) t] (n_{A}^2+ n_{AA}
n_{A})-n_{A} n_{AA}}
\end{equation}

In Fig.(2) the variation of susceptibility with respect to
temperature at a certain concentration value (c=1) for different q
values, in other words the plot of the function $\chi_{q}=f(T)$ is
represented. In this plot solid line correspond to $q\cong1$ and
the dashed line to $q=0.5$.

\begin{figure}[t]
\resizebox{0.45\textwidth}{!}{%
  \includegraphics{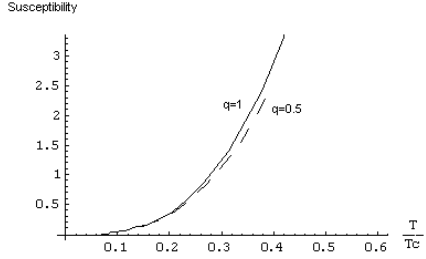}
}
\caption{Variation of susceptibility heat with respect to
temperature for simple cubic structure ($\gamma=6$).) at a certain
concentration ($c=1$) and for different g values (Eq(42)).
$\chi=f(T)$ plot.}
\label{fig:11}       
\end{figure}

When the magnetic part of the energy is taken under consideration
one writes:

\begin{equation}\label{Eq:42}
\frac{E_q}{N}=\frac{\gamma J}{4} (4 \eta-n_{AA})
\end{equation}

\begin{equation}\label{Eq:43}
\frac{E_q}{N}=\frac{-\gamma J}{4} n_{AA}
[\frac{1-e^{\frac{-J}{kT}}\sec(h_q) 2 (\gamma-1)
t}{1+e^{\frac{-J}{kT}}\sec(h_q) 2 (\gamma-1) t}]
\end{equation}

Using these expressions the specific heat is found to be;

\begin{equation}\label{}
    \frac{C_{q}}{k_{B}N}=\frac{1}{k_{B}N}\frac{dE_{q}}{dT}).
\end{equation}

\begin{equation}\label{Eq:44}
\frac{C_q}{kN}=\frac{\gamma J}{4 k} n_{AA} \frac{d}{dT}
[\frac{e^{\frac{-J}{kT}}\sec(h_q) 2 (\gamma-1)
t-1}{e^{\frac{-J}{kT}}\sec(h_q) 2 (\gamma-1) t+1}]
\end{equation}

\begin{figure}[t]
\resizebox{0.45\textwidth}{!}{%
  \includegraphics{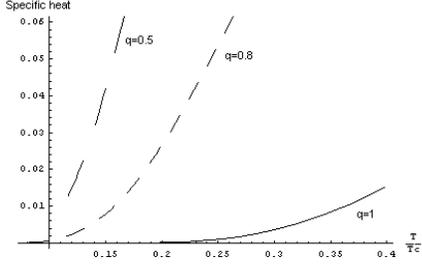}
}
\caption{Variation of specific heat with respect to temperature
for simple cubic structure ($\gamma=6$.) at a certain
concentration (c=1) and for different g values (Eq(44)). $SI=f(T)$
plot.}
\label{fig:11}       
\end{figure}

In Fig.(3) the variation of the specific heat with respect to
temperature at a certain value of concentration (c=1) for
different g values, in other words $c_{q}=f(T)$ plot is shown;
where the solid line is for $q\cong1$, closer spaced dashed line
for $q\cong0.8$ and wider spaced dashed line for $q=0.5$.

\section{Conclusions}
 In this study; by considering the dilute magnetic
systems as an ensemble of interacting elementary moments and in
this context with the help of spin-1 Ising model; they are
investigated microscopic level. Due to the statistical mechanics a
bridge has been constructed between microscopic level approach and
macroscopic experimental results. Magnetization is a long range
phenomenon with a memory. Such systems are non-extensive. For this
reason; magnetization has been considered in the frame of
non-extensive statistical mechanics. Starting with the standard
approach the generalization process proceeds. Same of the
experimental studies where this type of systems are investigated
has also been given [10,11,12]. In this study the variations of
magnetization, susceptibility and specific heat with respect to
temperature have been investigated. It is observed that; at
different values of the entropic index q, in other words when q is
decreasing, magnetization exhibits a linear variation rather than
a parabolic one. On the other hand, it seen that the
susceptibility does not undergo any change with q values. In the
specific heat however, with decreasing q temperature dependence
increases.

\end{document}